\documentclass[reprint,amssymb,pra,longbibliography]{revtex4-2}
\usepackage[utf8]{inputenc}
\usepackage[T1]{fontenc}
\usepackage{graphicx}
\usepackage{amsmath}
\usepackage{amssymb}
\usepackage{mathrsfs}
\usepackage{braket}
\usepackage{geometry}
\usepackage{here}
\usepackage{lmodern}
\usepackage{array}
\usepackage{natbib}
\usepackage{url}
\usepackage{textcase}
\usepackage{bm}
\usepackage{natbib}
\begin{document}
\title{Near-threshold scaling of resonant inelastic collisions at ultralow temperatures}
\author{Rebekah Hermsmeier$^{1}$, Adrien Devolder$^{2}$, {Paul Brumer$^{2}$, and Timur V. Tscherbul$^{1}$}}

\affiliation{$^{1}$Department of Physics, University of Nevada, Reno, NV, 89557, USA\\
$^{2}$Chemical Physics Theory Group, Department of Chemistry, and Center for Quantum Information and Quantum Control, University of Toronto, Toronto, Ontario, M5S 3H6, Canada}

\begin{abstract}
We show that the cross sections for a broad range of resonant  {\it inelastic} processes accompanied by excitation exchange (such as spin-exchange, F\"orster resonant, or angular momentum exchange)  exhibit an unconventional near-threshold scaling $E^{\Delta m_{12}}$, where $E$ is the collision energy, $\Delta m_{12}=m_1'+m_2'-m_1-m_2$, and $m_i$ and $m_i'$ are the initial and final angular momentum projections of the colliding species ($i=1,\,2$). In particular, the inelastic cross sections for  $\Delta m_{12}=0$  transitions display an unconventional $E^0$ scaling similar to that of elastic cross sections,  and  their rates vanish as $T^{\Delta m_{12}+1/2}$. For collisions  dominated by even partial waves (such as those of identical bosons in the same internal state) the scaling is modified to $\sigma_\text{inel}\propto E^{\Delta m_{12} +1} $ if $\Delta m_{12}$ is odd.  
 We present accurate quantum scattering calculations that illustrate these modified threshold laws for resonant spin exchange in ultracold Rb~+~Rb and O$_2$~+~O$_2$ collisions. Our results illustrate  that the threshold  scaling of collision cross sections is determined only by the energetics of the underlying process (resonant vs. exothermic) rather than by whether the internal states of colliding particles is changed in the collision.
 


\end{abstract}

\date{\today}
	\maketitle
{\it Introduction.} The unique controllability of ultracold atomic and molecular collisions \cite{Book_Krems,Chin2010,Carr:09,Balakrishnan:16,Bohn:17,Wolf:17,Matsuda:20,Yan:20,Liu:20,Devolder:21} gives rise to numerous applications of ultracold quantum gases in quantum information science and on precision tests of foundations of chemical and statistical physics \cite{Bohn:17,Liu:21}.
Further, much progress has been achieved over the last few years in our ability to observe the reactants and products of  molecular collisions and chemical reactions  in single, well-controlled quantum states \cite{Wolf:17,Matsuda:20,Yan:20,Liu:20,Liu:21}. Recent examples include precision studies of photoexcited reaction complexes \cite{Liu:20} and product state distributions \cite{Liu:21} in the ultracold chemical reaction KRb~+~KRb $\to$ K$_2$ + Rb$_2$,
 electric field-induced dipolar exchange resonances in KRb~+~KRb collisions \cite{Matsuda:20}, and microwave shielding of NaK~+~NaK collisions \cite{Yan:20}. In addition, observations have been made of ultracold atom-molecule collisions  in an optically trapped Na~-~NaLi($^3\Sigma$) mixture, of CaF$(^2\Sigma)$~+~CaF$(^2\Sigma)$  collisions in an optical dipole trap  \cite{Cheuk:20} and of cold O$_2(^3\Sigma)$~+~O$_2(^3\Sigma)$ collisions in a magnetically trapped gas of oxygen molecules  \cite{Segev:19}. 
 

 
 Ultracold atoms and molecules are typically prepared and trapped in quantum states with a fixed value of the angular momentum projection $m$ on a space-fixed quantization axis (typically fixed by an external electromagnetic field). 
 Angular momentum projection-changing collisions  play a key role in ultracold  chemistry. If such collisions are exothermic, they convert  the intramolecular energy into  translational energy of the relative motion,  leading to undesirable trap loss   and heating. In contrast, resonant excitation exchange (EE) collisions conserve the internal energy but lead to a coherent transfer of excitations (such as spin polarization) from one particle to another.  Examples of resonant EE processes  include spin-exchange in atomic, molecular, and ionic collisions \cite{Walker:97,Walker:11,Tscherbul:09,Tscherbul:11,Savukov:05,Sikorsky:18b,Rui:17,Liu:19,Kawaguchi2012,StamperKurn:13,Nishida:13,Bauer:13},  F{\"o}rster  resonances in collisions of Rydberg atoms \cite{Safinya:81,Zanon2002,Ryabtsev2009,Nipper2011,Ravets:14,Win2018},
atom-dimer exchange chemical reactions  \cite{Sikorsky:18,Rui:17}, excitation exchange between identical atoms or molecules \cite{Bouledroua2001}, charge transfer in cold ion-atom collisions \cite{Cote2000,Bodo:08,Zhang2009}, and rotational angular momentum projection-changing collisions \cite{Devolder:21}.
 
   Spin-exchange collisions, in particular, play an important role in a wide array of research fields, ranging from spin-exchange optical pumping \cite{Walker:97,Walker:11,Tscherbul:09,Tscherbul:11}, cold chemistry \cite{Sikorsky:18,Rui:17,Liu:19}, precision magnetometry \cite{Savukov:05}, and astrophysics \cite{Zygelman:05,Zygelman:10,Ming:15,Glassgold:05,Furlanetto:07} to quantum many-body physics \cite{Kawaguchi2012,StamperKurn:13,Nishida:13,Bauer:13} and quantum information processing, where they are used to generate entangled states \cite{Duan:02,Chang:05} and to operate quantum logic gates \cite{Jaksch:99,Calarco:00,Hayes:07,Jensen:19,Ni:18}.
 Recent experiments observed spin-exchange collisions of ultracold alkali-metal atoms \cite{Schmidt:19,Fang:20}, F\"orster resonant energy transfer in Rydberg atom collisions \cite{Ryabtsev2009,Nipper2011,Ravets:14}, and electric field-induced resonances in collisions of rotationally excited KRb molecules \cite{Matsuda:20}. We have recently found that ultracold EE collisions are amenable to extensive coherent control  \cite{Devolder:21}  due to the suppression of $m_{12}$-changing collisions at ultralow temperatures.
 
 However, despite the ever-increasing  interest in ultracold EE collisions, there remains a considerable uncertainty about the scaling of their integral cross sections (ICS) near collision thresholds \cite{Wigner:48,Sadeghpour:00}. It is commonly believed that  the inelastic ICSs scale as $k^{-1}$  with the incident collision wavevector $k$ in the limit $k\to 0$ \cite{Balakrishnan:97,Krems:05}.  While some computational studies did observe the $k^{-1}$ scaling \cite{Tscherbul:09,Ming:15,Liu:19} others reported that the ICS for atomic EE collisions approach a constant value  \cite{Glassgold:05,Furlanetto:07,Bodo:08,Zygelman:10}, a highly unexpected observation that has, to our knowledge, remained unexplained. Several authors attributed the $k^0$ scaling to a deficiency of the elastic approximation \cite{Dalgarno:61,Dalgarno:65,Moerdijk:96}, which effectively neglects the hyperfine structure of colliding atoms \cite{Zygelman:10,Ming:15,Sikorsky:18b}. More recently, the constant near-threshold scaling of spin-exchange cross sections was claimed to be incorrect \cite{Ming:15}, further deepening the existing controversy surrounding the threshold behavior of EE collisions.
 
  Krems and Dalgarno presented a derivation of the threshold laws for $m$-changing collisions of an open-shell atom (or rotating molecule)  with a spherically symmetric particle \cite{Krems:03}. However, useful as it is, their analysis could not be applied to describe EE processes, where both species exchange their angular momenta. As such, a comprehensive understanding of the threshold behavior of EE processes is still lacking, limiting our  ability to control a wide range of ultracold EE collisions that are key to implementing quantum logic gates \cite{Jaksch:99,Calarco:00,Hayes:07,Jensen:19,Ni:18} and to generating entanglement \cite{Duan:02,Chang:05} and strong correlations  \cite{Kawaguchi2012,StamperKurn:13,Bauer:13,Nishida:13} in quantum many-body systems.


Here, we resolve the long-standing controversy regarding the threshold behavior of EE cross sections. By deriving the threshold laws
 for ultracold EE collisions of two species {\it both of which possess internal angular momenta}, we show that the ICS for these processes display a collision energy scaling {$\sigma_\text{EE} \propto E^{|\Delta m_{12}|}$. Here, $\Delta m_{12}=m_1'+m_2'-m_1-m_2$, and $m_i$ and $m_i'$ are the initial and final angular momentum projections of the colliding species ($i=1,\,2$). In the important particular case $\Delta m_{12}=0$, the $s$-wave {\it inelastic} ICS approaches a constant value and the corresponding rate vanishes as $T^{1/2}$ as typically expected of elastic collisions \cite{Krems:05}}. To our knowledge, these results provide the first conclusive explanation for the unconventional $k^0$ near-threshold scaling of the resonant spin-exchange and charge exchange processes observed in a number of previous quantum scattering calculations \cite{Zygelman:10,Bodo:08}.
To illustrate our findings, we present numerically exact quantum scattering calculations of ultracold O$_2$~+~O$_2$ and Rb~+~Rb collisions, which confirm our analytic results  and provide a blueprint for observing the unconventional threshold scaling of inelastic cross sections in ultracold atomic and molecular collision experiments. (For clarification, note that inelastic refers to any collision where the nature or internal state of a colliding partner changes \cite{CohenTannoudji,Weiner:99}).


{\it Theory.} Consider a binary collision of two atoms and/or molecules, each possessing {\it total} internal angular momenta  $\mathbf{j}_1$ and $\mathbf{j}_2$. The nature of the operators $\mathbf{j}_i$ is set by the details of the internal structure of the colliding species, which   {is immaterial for the discussion below}. We focus on the case of weak external fields,   {where}  $j_i$ remain good quantum numbers, and consider as numerical examples, collisions of two open-shell $^{17}$O$_2(^3\Sigma)$ molecules ($\mathbf{j}=\mathbf{S}+\mathbf{N}$, where $S$ is the electron spin and $N$ is the rotational angular momentum)   {and} of two   {$^{87}$Rb} atoms ($\mathbf{j}=\mathbf{S}+\mathbf{I}$, where $\mathbf{I}$ is the atomic nuclear spin).  

The starting point for our discussion is the expression for the integral cross section for two particles initially colliding in well-defined angular momentum eigenstates $|j_i m_i\rangle$ (the subscripts $j_1$ and $j_2$ will be omitted for brevity unless stated otherwise)
\begin{multline}\label{ICS}
\sigma_{m_1 m_2\to m_1' m_2'}=\frac{\pi}{k^2} \sum_{l, m_l} \sum_{l’,m’_l}  |T_{m_1 m_2,lm_l\to m_1' m_2',l’m’_l}|^2
\end{multline}
{where $|lm_l\rangle$ and $|l'm_l'\rangle$ are the eigenstates of the orbital angular momentum $\hat{L}^2$ of the collision complex and its projection $\hat{L}_z$ on the space-fixed quantization axis.}
We are interested in the threshold scaling of the ICS with the collision energy $E$, which is determined by that of the transition $T$-matrix elements $T_{m_1 m_2, lm_l\to m_1' m_2',l’m’_l}$. To make the threshold scaling more explicit, we rewrite  the $T$-matrix  in the total angular momentum representation \cite{Arthurs:60,Rowe:79} 
\begin{multline}\label{T_totalJrep}
T_{j_1m_1 j_2m_2,l m_l\to j_1'm_1'j_2'm_2',l’m’_l}
=\sum_{J,M} \sum_{j_{12},m_{12}} \sum_{j’_{12},m’_{12}} (2J+1) \\
 \sqrt{(2j_{12}+1)(2j’_{12}+1)} (-1)^{l’+l} T^J_{j_1 j_2 j_{12}l\to j_1' j_2' j’_{12} l’}\\
(-1)^{j_{12}+j’_{12}+m_{12}+m’_{12}}
\begin{pmatrix}
j_1&j_2&j_{12}\\
m_1&m_2&-m_{12}
\end{pmatrix}
 \begin{pmatrix}
 j_{12}&l&J\\
 m_{12}&m_l&-M
\end{pmatrix}\\
(-1)^{j_1+j’_1-j_2-j’_2}
\begin{pmatrix}
j’_1&j’_2&j’_{12}\\
m’_1&m’_2&-m’_{12}
\end{pmatrix} 
\begin{pmatrix}
j’_{12}&l’&J\\
 m’_{12}&m’_l&-M
 \end{pmatrix}
\end{multline}
where $j_{12}=|\mathbf{j}_{12}| = |\mathbf{j}_{1} + \mathbf{j}_{2}|$ is the total internal angular momentum, and   $J = |\mathbf{J}| = |\mathbf{j}_1 + \mathbf{j}_2 + \mathbf{l}|$ is the total angular momentum of the collision complex, which is conserved in the absence of external fields. 

The advantages of using Eq. (\ref{T_totalJrep}) are two-fold.
{First,  the threshold behavior of the $T$-matrix elements $T^J_{j_1 j_2 j_{12}l\to j_1' j_2' j’_{12} l’}$ only depends on $l$ and $l'$ through the Wigner threshold law} \cite{Wigner:48,Sadeghpour:00}
 \begin{equation}\label{TmatrixWignerGeneral}
T_{\gamma l,\gamma'l'} \propto k^{l+1/2}(k')^{l'+1/2}, 
\end{equation}
 where $k$ and $k'$ are the incident and final  wavevectors \cite{Wigner:48,Ross:61,Sadeghpour:00} and $\gamma$ stands for the internal quantum numbers $j_1$, $j_2$, and $j_{12}$ (i.e. all quantum numbers other than $l$). We note that Eq.~(\ref{TmatrixWignerGeneral}) assumes the absence of near-threshold resonance, bound, and virtual states. A different $k$ scaling would result if such states were present \cite{Sadeghpour:00,Ross:61,Simbotin:14}. In the case of resonant scattering in the absence of external fields considered below, the initial and final states are degenerate ($k=k'$) and the near-threshold dependence of $T$-matrix elements takes the form
 \begin{equation}\label{TmatrixWigner}
T_{\gamma l,\gamma'l'} \propto k^{l+l'+1}.
\end{equation}
In the limit of zero collision energy ($k\to 0$) it follows from Eq. (\ref{TmatrixWigner}) that the $T$-matrix elements with the lowest $l$ and $l’$ provide the dominant contributions to the sum in Eq.~(\ref{T_totalJrep}) and hence to the ICS (\ref{ICS}).

{Second, the 3$j$-symbols $(:::)$  in Eq. (\ref{T_totalJrep}) make explicit the rotational symmetry restrictions on the possible values of  $l,m_l,l'$, and $m'_l$.} The 3-$j$ symbols must satisfy the selection rule $m_{12}+m_l=M=m’_{12}+m’_1$,$m_1+m_2=m_{12}$, and $m’_{12}=m’_1+m’_2$, which implies conservation of the total angular momentum projection, $M=m_1+m_2+m_l=m_1'+m’_2+m’_l$. The values of $m_l$ and $m'_l$ are then restricted by the change of the internal projection $\Delta m_{12} = m_{12} - m_{12}'= m_l'-m_l$. To illustrate the restrictions on $l$ and $l'$, consider the $s$-wave  scattering ($l=m_l=0$) of two distinguishable particles that changes $m_{12}=m_1+m_2$ by 1. As stated above, since $M$ is conserved, we have $\Delta m_{12} = m_{12} - m_{12}'= m_l'-m_l = \pm 1$ and thus  $m_l'=\pm 1$. 
{In the absence of additional symmetry restrictions on the lowest possible values of $l$ or $l'$ in  Eq.~(\ref{ICS}),  the dominant partial wave contributions are $l=0$ and $l'=|\Delta {m_l}| = |\Delta m_{12}|$. Substituting these values into Eq.~(\ref{TmatrixWigner}) we find $T_{\gamma l,\gamma'l'} \propto k^{|\Delta m_{12}|+1}$.

For the special case of spin-exchange transitions, which only change the projections of the internal angular momenta  ($j_1=j'_1$ and $j_2=j'_2$), the conservation of the total    {inversion} parity $p=(-1)^{j_1+j_2+l}$
 implies that $l$ and $l'$ must have the same parity. 
For ultracold collisions $l=0$ and thus $l'$ must be even. The relation $l' =  |\Delta m_{12} |$ still holds for even $|\Delta m_{12}|$ while for odd values of $|\Delta m_{12}|$, it must be replaced by $l'= |\Delta m_{12}|+1$. 

 For collisions of identical bosons (fermions) in the same internal state, the permutation symmetry must also be taken into  account. Only even (odd) partial waves will be present in Eq.~(\ref{ICS}) \cite{Green:75,Tscherbul:09}. For ultracold collisions of identical bosons $l=0$, and all permissible values of $l'$ are even as the case of    {$j_i$-conserving} spin-exchange transitions. {For ultracold collisions of identical fermions, $l=1$ and all permissible values of $l'$ are odd. The relation $l' =  |\Delta m_{12} |$ holds for odd $|\Delta m_{12}|$ while for even values of $|\Delta m_{12}|$, it must be replaced by $l'= |\Delta m_{12}|-1$. Again, we find the same near-threshold scaling $T_{\gamma l,\gamma'l'} \propto k^{|\Delta m_{12}|+1}$ for even $|\Delta m_{12}|$ and $T_{\gamma l,\gamma'l'} \propto k^{|\Delta m_{12}|+2}$ for odd $|\Delta m_{12}|$. Note that $|\Delta m_{12}|=0$ is a special case: the lowest value of $l'$ is 1 and $T_{\gamma l,\gamma'l'} \propto k^3$.  Thus, ultracold inelastic collisions of identical fermions with $|\Delta m_{12}|=0,1$ and $2$ all follow a near-threshold scaling identical to that of $p$-wave elastic scattering, i.e., $T_{\gamma l,\gamma'l'} \propto k^3$.}

Summarizing the above discussion, the threshold behavior of the $T$-matrix elements    {in the absence of $l$-restricting symmetries} is given by $T_{\gamma l,\gamma'l'} \propto k^{|\Delta m_{12}|+1}$. By taking the absolute magnitude squared of the $T$-matrix elements and dividing by $k^2=2\mu E$, where $\mu$ is the reduced mass of the collision complex [see Eq.~(\ref{ICS})], we obtain the threshold scaling of the ICS  for collision-induced EE processes in the general case 
\begin{align}
\sigma_{j_1 m_1 j_2 m_2 \to j_1' m_1' j_2' m_2'} &\simeq E^{|\Delta m_{12}|}.
\end{align}
A modified threshold scaling applies in the case of even partial wave scattering (spin-exchange transitions with $j_1=j_1'$ and $j_2=j_2'$ or those between identical bosons in the same internal states)
\begin{align}\label{sigma_Symmetry}\notag
\sigma_{j_1 m_1 j_2 m_2 \to j_1' m_1' j_2' m_2'} &\simeq E^{|\Delta m_{12}|} \quad\, (|\Delta m_{12}|\,\, \text{even}),\\
&\simeq E^{|\Delta m_{12}|+1} \, (|\Delta m_{12}|\,\, \text{odd}).
\end{align}
Significantly, unlike in the simplified case considered by Krems and Dalgarno \cite{Krems:03},  $m_i$ can be nonzero for {\it both} of the colliding species, and thus {\it $|\Delta m_{12}|=0$ can correspond to inelastic as well as to elastic scattering}. Therefore, remarkably, these expressions show that the  threshold scaling of the ICS for an inelastic EE process that conserves $m_{12}$ (such as flip-flop spin-exchange collisions) {\it approach a constant} value in the limit of ultralow collision energies. This behavior stands in contrast with the expected $E^{-1/2}$ scaling  of the inelastic ICSs in the limit $E\to 0$ \cite{Krems:05,Balakrishnan:16}, and was reported without explanation in several previous quantum scattering calculations of near-resonant charge exchange  \cite{Bodo:08} and spin-exchange in cold H~+~H collisions \cite{Zygelman:10}. 

 The physical origin of the unconventional Wigner threshold scaling can be traced back to the threshold behavior of the $T$-matrix elements, which depends {\it only on the initial  and final  values of  collision wavevectors $k$ and $k'$ and on $l$ and $l'$} (see above and Eq.~(\ref{TmatrixWignerGeneral}) \cite{Wigner:48}). As a result, as long as a collision process remains resonant ($k=k'$)  it will be governed by the same threshold dependence, regardless of whether the process is elastic or inelastic. Thus, as shown above, both the elastic and inelastic collisions that conserve $k$ and $\Delta m_{12}$ have the same constant near-threshold scaling,  usually characteristic of the elastic ICS.

Note that the above discussion can be straightforwardly extended  to exothermic (relaxation) collisions. Assuming that $k'$ is outside of the threshold regime,  the dependence of the $T$-matrix element on $k'$ is negligibly weak, and $T_{\gamma l,\gamma'l' } \propto k^{l+1/2}$. As a result, the relaxation ICSs assume their conventional $k^{-1}$ scaling with relaxation rates approaching constant values in the $T\to 0$ limit.

 

{\it Numerical examples: Spin exchange in ultracold O$_2$~+~O$_2$ and Rb~+~Rb collisions.}
To verify the unconventional Wigner threshold scaling of  inelastic EE processes derived above, we 
carried out numerically exact quantum scattering calculations of spin exchange in ultracold  O$_2$~+~O$_2$ and Rb~+~Rb collisions.

We first consider ultracold collisions of identical O$_2(X^3\Sigma^-)$ molecules initially in their ground rovibrational states $|N_i=0,j_i=1,m_i\rangle$, where  $\mathbf{j}_i=\mathbf{N}_i+\mathbf{S}_i$ ($i=1,2$) and $m_i$ is the eigenvalue of ${j}_{i_z}$. Since all $j_i>1$ states are energetically unavailable at ultralow collision energies, the initial and final two-molecule states can be denoted $|m_1 m_2\rangle$ with $m_i=-1,0,1$ (here, we  neglect the molecular hyperfine structure for simplicity). Additionally, the initial states for identical bosons (such as $^{17}$O$_2$)  must be symmetrized to account for identical particle permutation symmetry \cite{Tscherbul:09}
	\begin{equation}
	\ket{{m_1,m_2}}=\frac{1}{\sqrt{2(1+\delta_{m_1,m_2})}}\left[\ket{m_1}\ket{m_2}+\ket{m_2}\ket{m_1} \right].
	\end{equation} 
	where $m_2\ge m_1$.

Figure~\ref{fig:O2O2}(a) shows the ICS for spin exchange in  $^{17}$O$_2$~+~$^{17}$O$_2$ collisions  calculated as a function of collision energy   {by integrating the coupled-channel (CC) Schr\"odinger equations  \cite{Tscherbul2008} on the radial grid  extending out to $R_\text{max}=200 a_0$. The other computation parameters are identical   to  those employed in Ref.~\cite{Devolder:21}.}
The ICS for the $|\Delta m_{12}|$ conserving $|-1,+1\rangle \to |00\rangle$ transition  is seen to approach a constant value in the limit  $E\to0$ as predicted  by Eq.~(\ref{sigma_Symmetry}). The corresponding inelastic rate tends to zero as $T^{1/2}$ as illustrated in Fig.~\ref{fig:O2O2}(b). Note that for $j=1$, the only transition with $|\Delta m_{12}|=0$ is between the states $\ket{{0,0}}$ and $\ket{{-1,+1}}$.

In Fig.~\ref{fig:O2O2}(a) we plot the ICS for $|\Delta m_{12}|$ changing transitions  
for two identical bosons in the same internal state ($\ket{0,0}\rightarrow \ket{+1,+1}$ and $\ket{0,0}\rightarrow \ket{0,+1}$),  and for  two identical bosons in different internal states ($\ket{-1,+1}\rightarrow \ket{0,+1}$). As follows  from Eq.~(\ref{sigma_Symmetry}), these three transitions must have the same $E^2$ scaling, as is indeed observed. The corresponding inelastic rate tends to zero as $T^{5/2}$ as illustrated by the fit in Fig.~\ref{fig:O2O2}(b).

\begin{figure}[t]
	\centering
	\includegraphics[width=0.96\columnwidth, trim = 20 105 30 35]{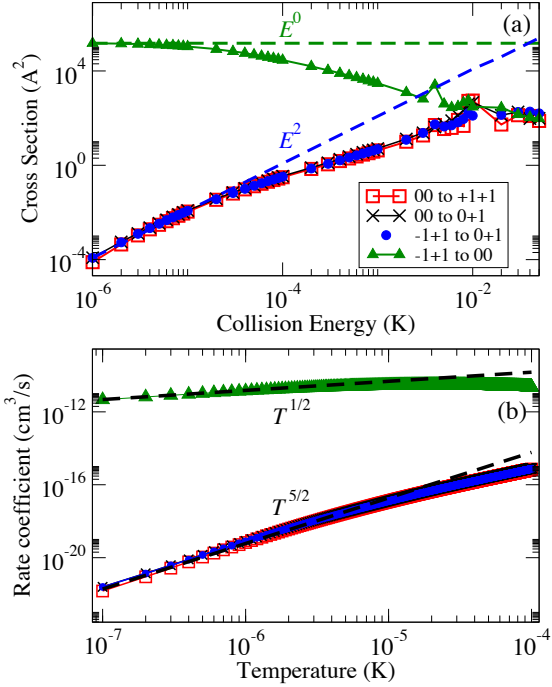}
	\caption{(a) State-to-state inelastic ICSs for ultracold O$_2$~+~O$_2$  collisions plotted as a function of collision energy.  The  initial and final values of $m_i$ for the O$_2$ molecules  are indicated in the  legend as $m_1m_2$ to $m_1'm_2'$. The fits as $E^0$ and $E^2$ are represented by dashed lines.  (b) Same as panel (a) but for the temperature dependence of state-to-state inelastic collision rates $K_{m_1m_2\to m_1'm_2'}$. The fits as $T^{1/2}$ and $T^{5/2}$ are represented by dashed lines. }
	\label{fig:O2O2}
\end{figure}



 We next consider ultracold spin-exchange collisions of $^{87}$Rb atoms, which have been the subject of much experimental study (for a review, see \cite{Chin2010}).
  The two-atom threshold states $|j_1m_{1}\rangle|j_2 m_{2}\rangle = |F_1m_{F_1}\rangle|F_2 m_{F_2}\rangle$, where $|j_i m_{F_i}\rangle = F_i m_{F_i}\rangle$ are the atomic hyperfine states and $\mathbf{j}_i=\mathbf{S}_i+\mathbf{I}_i$ are the total atomic angular momenta, which are vector sums of the electron and nuclear spins of the $i$-th atom ($i=1,2$). The quantum scattering problem for  $^{87}$Rb~+~$^{87}$Rb is solved  using the standard CC   approach as described in, e.g., Ref.~\cite{Li:08} by integrating the CC equations on a grid of $R$ values from 2 to 400$a_0$ with a step size of $5 \times 10^{-3} a_0$.

  The  $s$-wave ICSs for $^{87}$Rb~+~$^{87}$Rb spin-exchange collisions are plotted as a function  of collision energy 
    in Fig.~\ref{fig:RbRb}(a) for three representative hyperfine transitions  $|10\rangle|10\rangle \to |1,-1\rangle|1,+1\rangle$, $|20\rangle|20\rangle \to |2,-1\rangle|2,+1\rangle$, and $|20\rangle|20\rangle \to |2,-2\rangle|2,+2\rangle$.
    As all of these transitions have $\Delta m_{12}=0$, Eq.~(\ref{sigma_Symmetry}) establishes that their ICSs should scale as $E^0$ in the  limit of zero collision energy, and their rates should decrease as $T^{1/2}$ as $T\to 0$. 
\noindent The predicted trends are clearly observable in Fig.~\ref{fig:RbRb}.

 \begin{figure}[t]
 	\centering
 	\includegraphics[width=0.97\columnwidth, trim = 20 110 30 35]{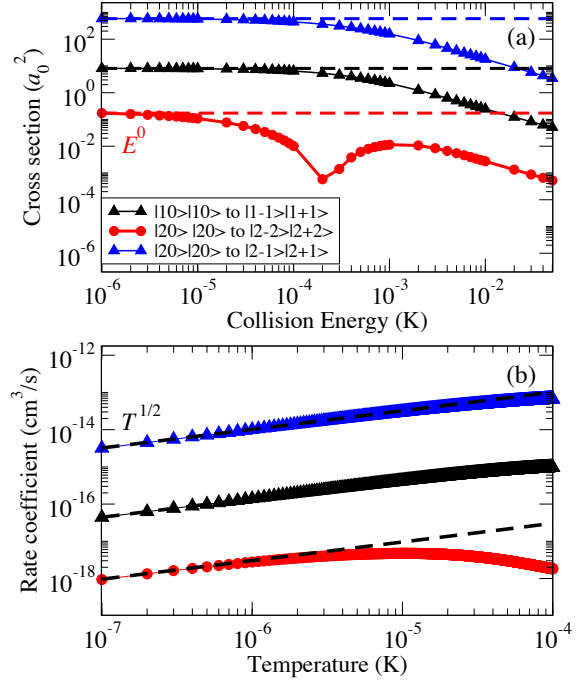}
 	\caption{(a) State-to-state inelastic ICSs for ultracold  $s$-wave $^{87}$Rb~+~$^{87}$Rb collisions plotted vs. collision energy at a magnetic field of 0.01 G.  The  initial and final hyperfine states of the Rb atoms $F_i m_{F_i}$ are indicated in the legend.The fits as $E^0$ are represented by green dots.
 		(b) Same as panel (a) but for the state-to-state inelastic collision rates $K_{m_{F_1} m_{F_2}\to m_{F_1}' m_{F_2}'}$ as a function of temperature. The fits as $T^{1/2}$ are represented by dashed lines.}
 	\label{fig:RbRb}
 \end{figure}

In conclusion, we have shown that the near-threshold scaling of the ICS for resonant inelastic EE processes (such as spin exchange)  is given by $\sigma_\text{inel} \propto E^{\Delta m_{12}}$ and only depends  on the difference between the combined angular momentum projections  $\Delta m_{12}$ in the incident and final collision channels. For $\Delta m_{12}=0$ the scaling of the inelastic ICS is the same as that of the elastic ICS, i.e., $\sigma_\text{inel} \propto E^0$. We have presented rigorous quantum scattering calculations to demonstrate the scaling for ultracold O$_2$~+~O$_2$ collisions in the experimentally accessible regimes.

Our work resolves the long-standing controversy concerning the threshold behavior of EE cross sections and confirms the correctness of the calculations reported in Refs.~\cite{Glassgold:05,Furlanetto:07}. More generally, it suggests that it is the energetics of a collisional EE process (resonant vs. exothermic) rather than the change in the internal states of collision partners, that determines the process' threshold behavior. 
Our results demonstrate a  universal   $T^{\Delta m_{12}+1/2}$  suppression of a wide class of resonant EE processes at ultralow temperatures, which could be observed experimentally for, e.g., spin-exchange atom-atom collisions in optical tweezers \cite{Schmidt:19}, atom-ion collisions in an optical lattice setup, which allows for high collision energy resolution  \cite{Ben-shlomi:21}, as well as in cold and ultracold collisions of Yb atoms \cite{Uetake:12}, Ti atoms \cite{Lu:09}, Rydberg atoms \cite{Ryabtsev2009,Nipper2011} and polar molecules \cite{Matsuda:20,Liu:21}.

We are grateful to Bernard Zygelman for bringing Ref.~\cite{Zygelman:10} to our attention.
This work was supported by the U.S. Air Force Office for Scientific Research (AFOSR) under Contract No.~
FA9550-19-1-0312 and partially by the NSF (Grant No. PHY-1912668). 


\bibliography{Ultracold_coherent_control}

\end{document}